\RequirePackage[l2tabu, orthodox]{nag}

\documentclass[10pt]{article} 

\usepackage{amsmath} 
\usepackage[a4paper]{geometry}
\usepackage{graphicx} 
\usepackage{microtype} 
\usepackage{siunitx} 
\usepackage{booktabs} 
\usepackage[colorlinks=false, pdfborder={0 0 0}]{hyperref} 
\usepackage{cleveref}

\usepackage[utf8]{inputenc} % cp1250, latin2
\usepackage[T1]{fontenc} 
\usepackage{indentfirst} 
\usepackage{amssymb} 
\usepackage{amsthm} 
\usepackage{algorithm} 
\usepackage{algpseudocode} 
\usepackage{color}
\usepackage{enumerate}
\usepackage{todonotes}
\usepackage{cite} 
\usepackage{needspace}

\usepackage{tikz}
\usetikzlibrary{positioning,chains,fit,shapes,calc,arrows,matrix} 
\tikzstyle{block} = [draw,rectangle,thick,minimum height=2em,minimum width=2em]

\newtheorem{theorem}{Theorem}

\newtheorem{example}{Example}

\newcommand\bb[1]{\textbf{#1}}

\title{Robust Approach to Restricted Items Selection Problem} 
\author{Maciej Drwal\footnote{Department of Computer Science, Wroclaw University of Science and Technology, Poland}}
\date{\today}

\begin{document}

\maketitle

\begin{abstract}
    We consider the robust version of items selection problem, in which the goal is to choose representatives from a family of sets, preserving constraints on the allowed items' combinations. We prove NP-hardness of the deterministic version, and establish polynomially solvable special cases. Next, we consider the robust version in which we aim at minimizing the maximum regret of the solution under interval parameter uncertainty. We show that this problem is hard for the second level of polynomial-time hierarchy. We develop an exact solution algorithm for the robust problem, based on cut generation, and present the results of computational experiments. 
    %\keywords{robust optimization \and combinatorial optimization \and mixed-integer programming}    
\end{abstract}

\section{Introduction}\label{sec:1}

In this paper we consider a robust variant of combinatorial optimization problem of selecting $p_i$ items from $m$ sets of items, $i=1,\ldots,m$, with the objective to minimize their total cost. We assume the possibility that some choices of pairs of items are forbidden. We further assume that costs of items are not given precisely, but only intervals of their true values are known.

This problem models many practical situations, when we need to design a feasible configuration, subject to uncertain costs of its elements. For example, consider designing a production line in a flexible manufacturing system
\cite{LEE200361, JAHROMI2012224}. The production line consists of $m$ sites arranged for performing a sequence of operations. Execution of each operation requires a specific machine tool, and thus for each operation the designer must choose a single eligible tool among a variety of alternatives (for example, belonging to different types or brands of tools, suitable for a given operation). The costs of tools are subject to random fluctuations, due to the number of factors, for example: time difference between design phase and the purchase phase, additional installation costs, additional customization costs, etc. Consequently, it is usually not possible to determine the exact cost of each tool at the design stage. Moreover, not every tool is compatible with each other, and it may not be possible to install arbitrary selection of tools in a common production line. Thus the choice of possible tool configurations is restricted. For decision maker, the goal is to select eligible tools for each operation's site, so that the total cost would be minimal. 

As another example, consider the workflow planning for an engineering project. Decision makers need to assign teams of people to work on developing components of the whole system. For each component they would list a group of candidates that are suitable to working on it. Then they would decide on the actual team members by selecting a required number of people from each group of candidates. Some people, being adequate candidates for more than one team, may be listed more than once. However, due to the assumed work schedule, they cannot be assigned to two teams that work simultaneously. The final cost of completing each component is uncertain, as it depends not only on the skills and experience of selected people, but also on components' complexity, applied technologies, and possibly many other factors.

\subsection{Related Work}

The basic variant of this problem has been first considered in \cite{dolgui2012min} under the name Representatives Selection Problem, where we are allowed to select one item from each set of alternatives. In order to alleviate the effects of cost uncertainty on decision making, the min-max and min-max regret criteria \cite{aissi2009min, kasperski2008discrete} have been proposed to assess the solution quality. The problem formulations using these criteria belong to the class of {\em robust} optimization problems \cite{roy2010robustness}. Such approach appears to be more suitable for large scale design projects than an alternative stochastic optimization approach \cite{kall1994stochastic}, when: 1)~decision makers do not have sufficient historical data for estimating probability distributions; 2)~there is a high factor of risk involved in one-shot decisions, and a precautionary approach is preferred. The robust approach to discrete optimization problems has been applied in many areas of industrial engineering, such as: scheduling and sequencing \cite{kasperskiminmax, drwal2016complexity, drwal2018robust}, network optimization \cite{bertsimas2003robust, averbakh2004interval}, assignment \cite{pereira2011exact, aissi2005complexity}, and others \cite{goerigk2016algorithm}.

Note that deterministic version of Representatives Selection Problem is easily solvable in polynomial time. For interval uncertainty representation of cost parameters the problem can still be solved in polynomial time, both in case of minimizing the maximum regret and the relative regret \cite{dolgui2012min}. However, in case of discrete set of scenarios, the problem becomes NP-hard even for 2 scenarios, and strongly NP-hard when the number of scenarios $K$ is a part of the input. In \cite{deineko2013complexity} authors prove that strong NP-hardness holds also when sets of eligible items are bounded. In \cite{kasperski2015approximability} an $O(\log K / \log \log K)$-approximation algorithm for this variant was given.

In this paper we consider a generalization of the Representatives Selection Problem, where it is required to select a specific number of items from each set, and constraints may be imposed on the selected items' configurations, by specifying the pairs of items that cannot be selected simultaneously. We assume the interval representation of cost uncertainty, since this case appears to be more interesting from both practical and theoretical point of view.

\section{Problem Formulation}

We start from defining a deterministic version of the considered problem, which we call Restricted Items Selection Problem (abbreviated \textsc{RIS}). Given are $m$ sets $I_i$ of items, with integer values $c_{ij}$ associated with each $j \in I_i$, $r_i = |I_i|$. Given is a set $T$ of tuples $(i,k,j,l)$, $i,j \in \{ 1,\ldots,m \}$, $k \in I_i$, $l \in I_j$, indicating that items $k$ and $l$ cannot be selected simultaneously (i.e., if $k$ is selected, then $l$ cannot be selected, and if $l$ is selected, then $k$ cannot be selected). The set $T$ contains all the {\em forbidden pairs}. Given are $m$ positive integers $p_i < |I_i|$. The goal is to select for each $i=1,\ldots,m$ a subset $S_i \subset I_i$ of exactly $p_i$ items, so that the total value of selected items is maximized. 

Let $x_{ij} = 1$ if $j$th item from $i$th set is selected, $x_{ij} = 0$ otherwise. The problem can be stated as: 
\begin{equation}\label{ris-obj} 
    \textrm{minimize } \sum_{i=1}^m \sum_{j=1}^{r_i} x_{ij} c_{ij} 
\end{equation}
subject to: 
\begin{align}
    \forall_i & \;\;\; \sum_{j=1}^{r_i} x_{ij} = p_i, \label{ris-1} \\
    \forall_{(i,k,j,l) \in T} & \;\;\; x_{ik} + x_{jl} \leq 1, \label{ris-2} \\
    \forall_{i,j} & \;\;\; x_{ij} \in \{ 0,1 \}. 
\label{ris-3} \end{align}

Let us now define the version of the problem with uncertain data, which we call Interval Min-Max Regret Restricted Items Selection Problem (abbreviated \textsc{IRIS}). For each item, given is an interval of possible costs, $c_{ij}^- \leq c_{ij} \leq c_{ij}^+$. The set of all cost scenarios is a product of intervals, denoted: 
\begin{align*}
    \mathcal{U} = \{ \bb{c}=(c_{11}, \ldots, c_{mr_m}) : c_{ij}^- \leq c_{ij} \leq c_{ij}^+, \\
    \; i=1,\ldots,m, \; j=1,\ldots,r_i \}. 
\end{align*}
Let $\bb{x}$ be the vector of variables $x_{ij}$ (indexing $(i,j)$ agreeing with the cost vector $\bb{c}$). We define the regret of a solution $\bb{x}$ as the difference between the solution value in the worst-case scenario, and the best possible value in the worst-case scenario: 
\begin{equation}\label{max-regret} 
    R(\bb{x}) = \max_{\bb{c} \in \mathcal{U}} \left( \bb{c}\cdot\bb{x} - \min_{\bb{y} \in F}\bb{c}\cdot\bb{y} \right), 
\end{equation}
where: 
\begin{equation}\label{F-constr} 
    \mathcal{F} = \{ \bb{x} : \; \forall_i \; \sum_{j=1}^{r_i} x_{ij} = p_i, \; \forall_{(i,k,j,l) \in T} \; x_{ik} + x_{jl} \leq 1, \; x_{ij} \in \{ 0, 1 \} \}
\end{equation}
is the set of feasible solutions (choices of items from each set $I_i$).
The objective of the \textsc{IRIS} problem is to find such a choice $\bb{x}$ that minimizes the maximum regret:
\begin{equation}\label{generic-obj} 
    \min_{\bb{x} \in \mathcal{F}} R(\bb{x}). 
\end{equation}

\subsection{Related Problems}

The considered problem generalizes two related fundamental combinatorial optimization problems, namely, the Selecting Items Problem \cite{averbakh2001complexity, conde2004improved}, and the aforementioned Representatives Selection Problem \cite{dolgui2012min}, of which the robust variants have been investigated in the literature. 
\begin{enumerate}
    
    \item \textsc{Selecting items problem (SI)}
    
    Given is a set $I$ of items, with a positive integer $c_i$, denoting item's cost, associated with each $i \in I$. Given is positive integer $p < |I|$. The goal is to select a subset $S \subset I$ of exactly $p$ items, so that the total value of $S$ is minimized. Let $\bb{x}$ be the characteristic vector of $S$, i.e., $x_{i} = 1$ if $i \in S$, and $x_i=0$ otherwise. The problem can be stated as: $$ \min \left\{ \sum_{i=1}^n x_i c_i : \; \sum_{i=1}^n x_i = p, \; \bb{x} \in \{ 0, 1 \}^n \right\}. $$
    
    \item \textsc{Representatives selection problem (RS)}
    
    Given are $m$ sets $I_i$ of items, with a positive integer costs $c_{ij}$ associated with each $j \in I_i$, $r_i = |I_i|$. The goal is to select a single item from each set, $J=(j_1, j_2, \ldots, j_m)$, $j_i \in I_i$, so that the total value is minimized. Let $x_{ij} = 1$ if $j$th item from $i$th set is selected, $x_{ij} = 0$ otherwise. The problem can be stated as: $$ \min \left\{ \sum_{i=1} \sum_{j=1}^{r_i} x_{ij} c_{ij} : \; \forall_i \; \sum_{j=1}^{r_i} x_{ij} = 1, \; \forall_{(i,j)} \; x_{ij} \in \{ 0,1 \} \right\}. $$ 
\end{enumerate}

If $p_i=1$ for each $i=1,\ldots,m$, and $T = \emptyset$, then problem \textsc{RIS} is equivalent to \textsc{RS}. If $T = \emptyset$, then problem \textsc{RIS} is equivalent to $m$ independent instances of \textsc{SI} problem.

\section{Problem Complexity}

Deterministic problems \textsc{SI} and \textsc{RS} can be solved in polynomial time (trivially using sorting). However, forbiddance constraints make these problems much more difficult in general.
\begin{theorem}\label{thm:1} 
    Problem \textsc{RIS} is NP-hard, even when $p_i = 1$ for $i=1,\ldots,m$. 
    %and $c_{ij} \in \{ 0, 1 \}$ for all $i=1,\dots,m$, $j \in I_i$.
\end{theorem}

{\em Proof.} We present reduction from the independent set problem \cite{garey2002computers}. Let $G=(V,E)$ be an undirected graph with $|V| = n$, and we ask if $G$ has an independent set of size at least $K$. We construct an instance of \textsc{RIS} as follows. Let us enumerate vertices in $V$ by consecutive numbers $1,\ldots,n$, denoting $i$th vertex by $v_i$. For each vertex $v_i$ we create an item set $I_i$, containing an item of value $0$. For each edge $(v_i, v_j) \in E$, we add tuple $(i,1, j,1)$ to set $T$, forbidding simultaneous choice of items corresponding to $v_i$ and $v_j$. To each set $I_i$ we also add a dummy item with value $1$.

Note that a feasible solution to \text{RIS} instance always exists (dummy items are not involved in constraints $T$). In such a solution we have a choice of items, one from each set $I_i$, such that the corresponding nodes $v_i$ are not directly connected in $G$. The cost of an optimal solution of \text{RIS} instance is less or equal to $n-K$ if and only if there is an independent set of size $K$ in $G$. \qed

\medskip

Consequently, the uncertain data problem \textsc{IRIS} is NP-hard as well. We can, however, identify the following special cases of \textsc{RIS} by imposing additional restrictions on the set of constraints. By exploiting these properties we are able to improve the solution algorithms of uncertain \textsc{IRIS} problem in these special cases (see Section \ref{sec:sol-alg}).
\begin{theorem}\label{thm:2} 
    Problem \textsc{RIS} can be solved in polynomial time in any of the following special cases: 
    \begin{enumerate}
        [1)] 
        \item Each item appears at most once in any constraint in $T$. 
        \item If $(i,k,j,l) \in T$ and $(j,l,p,q) \in T$, then $(i,k,p,q) \in T$ (that is, the ``forbiddance'' relation between pairs of items is transitive). 
    \end{enumerate}
\end{theorem}

{\em Proof.} To see that \textsc{RIS} can be solved in polynomial time in case 1), we observe that if each item is involved in at most one constraint in $T$, then the constraint matrix of integer linear program \eqref{ris-obj}--\eqref{ris-3} is totally unimodular. This can be verified using the following characterization (see \cite{schrijver1998theory}, chapter 19): a $(0,\pm 1)$ matrix with at most two nonzeros in each column is totally unimodual, if the set of rows can be partitioned into two classes, such that two nonzero entries in a column are in the same class of rows if they have different signs and in different classes of rows if they have the same sign. These conditions are satisfied if we take the rows corresponding to constraints \eqref{ris-1} as the first class, and rows corresponding to constraints \eqref{ris-2} as the second class.

To prove case 2), we construct an instance of min-cost max-flow network problem that is equivalent to the corresponding \textsc{RIS} instance. First, note that due to the transitivity of ``forbiddance'' relation, items can be grouped into equivalence classes. From each equivalence class, at most one item can be selected. If an item is not involved in any constraint, it would belong to an equivalence class of cardinality 1. The network consists of $m$ source nodes, two inner layers, followed by a single terminal node (see also Example \ref{example1} and Fig. \ref{fig:net1}). Each source node corresponds to the set $I_i$, $i=1,\ldots,m$, and provides a supply of $p_i$ units of flow. Nodes in the first inner layer correspond to items in sets $I_i$, and are connected by unit-capacity arcs to their corresponding source nodes. Each such arc has costs $c_{ij}$. The second inner layer consists of nodes corresponding to each equivalence class of items. Nodes from the first inner layer, that correspond to the items involved in a common equivalence class, are connected to the common node in the second layer. Arcs between first and second inner layer have unit-capacity and zero cost. Finally, there are unit-capacity and zero-cost arcs from each node in the second layer and the terminal node. The flow network can be constructed in polynomial time given an instance of \textsc{RIS}.  \qed

\begin{example}\label{example1}
    
    Consider an instance of \textsc{RIS} with $m=3$ item sets, each containing $r_i=3$ items, with the requirement to select $p_i=2$ items from each set. The set of forbidden pairs is $T = \{ (1,1, 2,1), (2,1,3,1), (1,1, 3,1), (1,2,2,2) \}$. Note that the first three pairs form a transitive subset, thus only one item can be selected from it. The corresponding flow network is presented on the Fig. \ref{fig:net1}. The first inner layer nodes correspond to items, denoted $i_{ij}$, and the second inner layer nodes, denoted $e_k$, correspond to the subsets of items from which only a single item can be selected.
    
\bigskip

\begin{figure}\label{fig:net1}
    \caption{Example of flow network illustrating case 2) of Theorem \ref{thm:2}.}
    \centering
\begin{tikzpicture}[auto]
    \small
    
    \foreach \pos/\name in {{(1,0)/p_3}, {(1,1)/p_2}, {(1,2)/p_1}}
            \node[circle,fill=black!33,inner sep=1pt] (\name) at \pos {$\name$};
            
    \foreach \pos/\lab/\name in {{(3,-1)/$i_{3,3}$/i33}, {(3,-0.5)/$i_{3,2}$/i32}, {(3,0)/$i_{3,1}$/i31},
                            {(3,0.5)/$i_{2,3}$/i23}, {(3,1)/$i_{2,2}$/i22}, {(3,1.5)/$i_{2,1}$/i21},
                            {(3,2)/$i_{1,3}$/i13}, {(3,2.5)/$i_{1,2}$/i12}, {(3,3)/$i_{1,1}$/i11}}
            \node[circle,fill=black!33,inner sep=0pt] (\name) at \pos {\lab};
            
    \foreach \pos/\name in {{(5,-1.5)/e_6}, {(5,-0.5)/e_5}, {(5,0.5)/e_4}, {(5,1.5)/e_3}, {(5,2.5)/e_2}, {(5,3.5)/e_1}}
            \node[circle,fill=black!33,inner sep=1pt] (\name) at \pos {$\name$};
            
    \node[circle,fill=black!33] (t) at (7,1) {t};
    
    \draw[->] (p_1) -- (i11);
    \draw[->] (p_1) -- (i12);
    \draw[->] (p_1) -- (i13);
    
    \draw[->] (p_2) -- (i21);
    \draw[->] (p_2) -- (i22);
    \draw[->] (p_2) -- (i23);
    
    \draw[->] (p_3) -- (i31);
    \draw[->] (p_3) -- (i32);
    \draw[->] (p_3) -- (i33);
        
    \draw[->] (i11) -- (e_1);
    \draw[->] (i21) -- (e_1);
    \draw[->] (i31) -- (e_1);
    
    \draw[->] (i12) -- (e_2);
    \draw[->] (i22) -- (e_2);
    
    \draw[->] (i13) -- (e_3);
    \draw[->] (i23) -- (e_4);
    \draw[->] (i32) -- (e_5);
    \draw[->] (i33) -- (e_6);
    
    \draw[->] (e_1) -- (t);
    \draw[->] (e_2) -- (t);
    \draw[->] (e_3) -- (t);
    \draw[->] (e_4) -- (t);
    \draw[->] (e_5) -- (t);
    \draw[->] (e_6) -- (t);
            
\end{tikzpicture}

\end{figure}
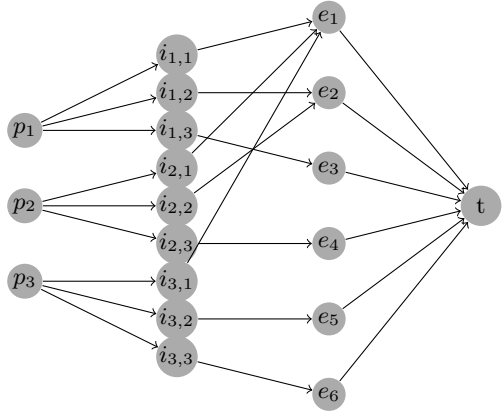

\end{example}

We conclude this section with the last statement regarding the complexity of general robust \textsc{IRIS} problem. It turns out that this problem is likely to be even much harder than its already NP-hard deterministic counterpart \textsc{RIS}. We show that it is hard for the complexity class $\Sigma_2^p = NP^{NP}$, a set of problems that remain NP-hard to solve even if we can use an oracle that answers NP queries in $O(1)$ time (i.e., the set of problems that can be solved in polynomial time on nondeterministic Turing machine that can access an oracle for NP-complete problem in every step of its computations) \cite{stockmeyer1976polynomial}.

The class $\Sigma_2^p$ can be also characterized as the set of all decision problems that can be stated in the second-order logic using a pair of universal and existential quantifiers. Let us define the following prototypical problem in this class:

\-\hspace{0.0cm} {\em Problem}: 2-Quantified 3-DNF Satisfiability (abbreviated $\exists\forall$ 3SAT)

\-\hspace{0.0cm} {\em Instance}: Two sets $X=\{ x_1, \ldots, x_s \}$, $Y=\{ y_1, \ldots, y_t \}$ of Boolean variables. A Boolean formula $\phi(X,Y)$ over $X \cup Y$ in disjunctive normal form, where every clause consists of exactly 3 literals.

\-\hspace{0.0cm} {\em Question}: Is $\exists_X \forall_Y \; \phi(X, Y)$ true? 
\begin{theorem}
    Problem $\exists\forall$ 3SAT is $\Sigma_2^p$-complete \cite{stockmeyer1976polynomial}. 
\end{theorem}

\begin{theorem}\label{thm:sigma2p}
    Problem \textsc{IRIS} with $p_i=1$, $i=1,\ldots,m$, is $\Sigma_2^p$-hard. 
\end{theorem}

{\em Proof.}
%($\Rightarrow$)
%Note that each clause is satisfied by exactly 1 assignment of truth variables that appear in it. Consequently, for all but a single assignment of its $x$-variables ($y$-variables), the clause necessarily evaluates to {\em false}.
Given any 3-DNF formula $\phi$ we construct and instance of \textsc{Interval Minmax Regret RIS} as follows. For each clause $\mathcal{C}_j$, $j=1,\ldots,m$, we create a set of items $I_j$. Each set contains three types of items: 

%\begin{enumerate}[1)]
1) an $X$-item that corresponds to the unique satisfying assignment of $x$-variables in clause $\mathcal{C}_j$,

2) a collection of $Y$-items, denoted $Y_{j,k}$, each corresponding to every assignment of $y$-variables that make the clause $\mathcal{C}_j$ evaluate to {\em false},

3) a special item.

%\end{enumerate}
We can assume that each clause contains at least one $y$-variable (otherwise the instance answer is always ``yes''). If there are no $x$-variables in a given clause, then there is no $X$-item in the set $I_j$ (only $Y$-items and a special item).

For each pair of item sets $I_i$, $I_j$, and for each pair of items $Y_{i,k} \in I_i$, $Y_{j,l} \in I_j$, we add a ``forbidding'' constraint $(i,k,j,l)$ to $T$, whenever:

1) both items correspond to assignment of common variable $y$, and,

2) item $Y_{i,k}$ corresponds to the assignment of variable $y=0$, while item $Y_{j,l}$ corresponds to the assignment $y=1$, or vice-versa.

Similarly, we add ``forbidding'' constraints for pairs of $X$-items from two sets $I_i$ and $I_j$, whenever their satisfying $x$-assignments contain a common Boolean variable, and they correspond to conflicting assignments of $x$-variables. Special items are not bound by any ``forbidding'' constraints.

For example, Let $\mathcal{C}_1 = (x_1 \wedge y_1 \wedge \bar{y}_2)$, and $\mathcal{C}_2 = (\bar{x}_1 \wedge y_1 \wedge y_3)$. Then $X$-item from set $I_1$ cannot be taken simultaneously with $X$-item from set $I_2$, since they both correspond to satisfying assignment of variable $x_1$, that is $x_1=1$ in $\mathcal{C}_1$, and $x_1=0$ in $\mathcal{C}_2$. Moreover there are 3 $Y$-items in $I_1$, corresponding to unsatisfying assignments of $(y_1,y_2)$, that is $00$, $01$ and $11$; and there are 3 $Y$-items in $I_2$, corresponding to unsatisfying assignments of $(y_1,y_3)$, that is $00$, $01$ and $10$. Note that since both clauses share variable $y_1$, items that correspond to a conflicting assignment form a ``forbidden pair'' (e.g., $Y$-item $00$ from the first set and $Y$-item $10$ from the second set, etc.).

Let $B > m$ be a large constant. For each set $I_j$, each $X$-item corresponding to the unique $x$-satisfying assignment has cost interval $[0, B]$. Each $Y$-item corresponding to unsatisfying assignment of $y$-variables has cost interval $[0, B^2]$. The special item has cost $c^- = c^+ = B+1$, regardless of the scenario. 

Note that in the worst-case scenario each item selected by the decision maker will have the interval upper-bound cost $c^+$, while all other items will have their interval lower-bound costs $c^-$.

Observe that any optimal solution of the considered \textsc{IRIS} problem corresponds to a valid assignment of truth values to all variables in $X \cup Y$: the choice of items encoded by the characteristic vector $\bb{x}$ corresponds to the assignment of $x$-variables, while the best alternative solution in the worst-case scenario, encoded by the characteristic vector $\bb{y}$ in \eqref{max-regret}, corresponds to the assignment of $y$-variables. Since $p_j=1$ for $j=1,\ldots,m$, from each set $I_j$ the decision maker selects a single item, deciding on the solution $\bb{x}$. Subsequently, the worst-case scenario is fixed, and the adversary selects a single item from each set $I_j$, deciding on an alternative solution $\bb{y}$.

The regret minimizing decision maker would always select either $X$-item or a special item from any set $I_j$. This follows from the fact that, if decision maker selects an $Y$-item, then in the worst-case scenario its cost will be $B^2$, while the adversary would always be able to choose an $X$-item with cost $0$. The decision maker would be better off never selecting an $Y$-item, having guaranteed cost no greater than $B+1$ (the cost of the special item, selected only when it is not possible to select $X$-item, due to ``forbidding'' constraints). On the other hand, the regret maximizing adversary would select one of $Y$-items, whenever possible, which would have cost $0$ in the worst-case scenario. Only when it is not possible to choose such an item (due to ``forbidding'' constraints) would the adversary select another item: either a special item, or an $X$-item, resulting in the cost $B$ or $B+1$. This happens when the adversary must choose an $y$-satisfying assignment in the corresponding clause.

Since the items must be selected preserving all the ``forbidding'' constraints, their choice corresponds to an assignment of truth values to all Boolean variables without any conflicts.

We will show that there exists an assignment of truth values to $x$-variables, such that for every assignment of $y$-variables at least one clause is satisfied, if and only if the value of a solution of the constructed instance is at most $Z = (m-1)B + m-1$.

Suppose that $\exists \forall$ 3SAT instance is positive (i.e., at least one clause would always be satisfied both by $x$- and $y$-variables). Then there exists an assignment of $x$-variables, consisting of $x$-satisfying assignment in at least $k$ clauses ($1 \leq k \leq m$), while in every $y$-assignment, at least one $x$-satisfied clause is also $y$-satisfied. In such case the total cost of decision maker's choice is $mB + (m-k)$ ($k$ $X$-items and $(m-k)$ special items). In at least one set $I_j$ the adversary cannot select $Y$-item, and thus is forced to select an $X$-item. But since that item is also selected by the decision maker, its cost is cancelled when evaluating the regret. Consequently, the total regret is no greater than $mB + (m-k) - B \leq (m-1)B + (m-1) = Z$.

Suppose now that $\exists \forall$ 3SAT instance is negative. Then for every assignment to $x$-variables consisting of $k$ $x$-satisfied clauses, the adversary can select an $y$-assignment such that all these $k$ clauses evaluate to {\em false}. Thus decision maker's cost is $mB + (m-k)$, while the adversary's cost is always zero; the adversary can always select either an $Y$-item with cost $0$, or, from a set $I_j$ that corresponds to a clause that is unsatisfied by $x$-variables, an $X$-item with cost $0$ (since in that case the decision maker must have taken a special item from $I_j$). Consequently, the maximum regret is at least equal to $mB > (m-1)B + (m-1) = Z$, which concludes the proof.

\qed
\begin{example}\label{example2}
    The following formula $\phi$ over $X = \{ x_1, x_2, x_3 \}$ and $Y = \{ y_1, y_2 \}$ is a positive instance of $\exists \forall$ 3SAT problem:
    $$ \boxed{ 
    \begin{array}{cccccccc}
         & \mathcal{C}_1 &  & \mathcal{C}_2 &  & \mathcal{C}_3 &  & \mathcal{C}_4 \\
         \phi = & (x_1 \wedge x_2 \wedge y_1) & \vee & (x_1 \wedge \bar{y}_1 \wedge y_2) & \vee & (x_2 \wedge \bar{x}_3 \wedge \bar{y}_2) & \vee & (x_3 \wedge \bar{y}_1 \wedge \bar{y}_2) 
    \end{array}
    } $$
    Consider the assignment of $x$-variables $(x_1, x_2, x_2) = (1,1,0)$. Then first 3 out of 4 clauses are $x$-satisfied, and it can be easily verified that all four possible $y$-assignments to $(y_1, y_2)$ would result in 1 out of the first 3 clauses also $y$-satisfied. This formula corresponds to the following instance of \textsc{IRIS} problem shown on Fig. \ref{fig:sets}.

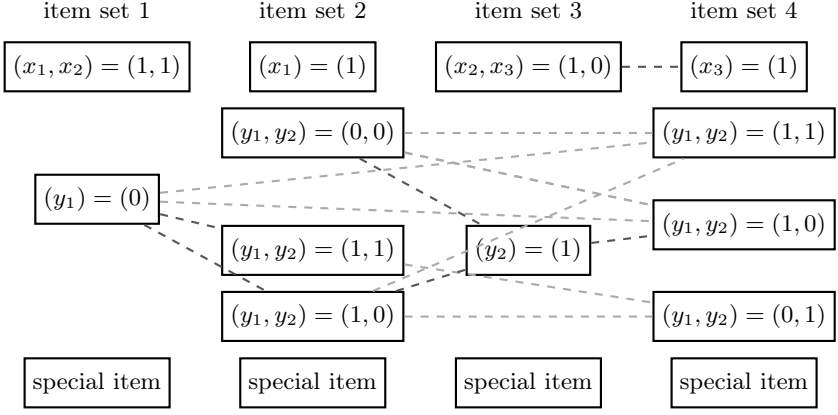
\begin{figure}\label{fig:sets}
    \caption{Illustration for Example \ref{example2}, showing the construction used in the proof of Theorem \ref{thm:sigma2p}. Each of the 4 sets contains: an $X$-item with cost interval $[0,B]$, a collection of $Y$-items with cost intervals $[0,B^2]$, and a special item with cost interval $[B+1,B+1]$. In the worst-case scenario, the items selected by decision maker have upper-bound costs, while all other items have lower bound costs.}
\begin{tikzpicture}[thick,scale=1, auto, >=stealth']
    \small
    \matrix[ampersand replacement=\&, row sep=0.2cm, column sep=0.4cm] {
        \node (I1) {item set 1}; \& \node (I2) {item set 2}; \& \node (I3) {item set 3}; \& \node (I4) {item set 4};\\
        
        \node[block] (X1) {$(x_1,x_2)=(1,1)$}; \& \node[block] (X2) {$(x_1)=(1)$}; \&
        \node[block] (X3) {$(x_2,x_3)=(1,0)$}; \& \node[block] (X4) {$(x_3)=(1)$}; \\
        
         \& \node[block] (Y21) {$(y_1,y_2)=(0,0)$}; \& 
         \& \node[block] (Y41) {$(y_1,y_2)=(1,1)$}; \\
        
        \node[block,anchor=south] (Y11) {$(y_1)=(0)$}; \& \node[block, anchor=north] (Y22) {$(y_1,y_2)=(1,1)$}; \& 
        \node[block,anchor=north] (Y31) {$(y_2)=(1)$}; \& \node[block] (Y42) {$(y_1,y_2)=(1,0)$}; \\
         
         \& \node[block] (Y23) {$(y_1,y_2)=(1,0)$}; \& \& \node[block] (Y43) {$(y_1,y_2)=(0,1)$}; \\
        
        \node[block] (S1) {special item}; \& \node[block] (S1) {special item}; \& 
        \node[block] (S1) {special item}; \& \node[block] (S1) {special item}; \\
    };

    \draw[dashed,draw=black!33] (Y11) -- (Y41);
    \draw[dashed,draw=black!33] (Y11) -- (Y42);   
    \draw[dashed,draw=black!33] (Y21) -- (Y41);
    \draw[dashed,draw=black!33] (Y21) -- (Y42);
    \draw[dashed,draw=black!33] (Y21) -- (Y42);
    \draw[dashed,draw=black!33] (Y22) -- (Y43);
    \draw[dashed,draw=black!33] (Y23) -- (Y41);
    \draw[dashed,draw=black!33] (Y23) -- (Y43);
    
    \draw[dashed,draw=black!66] (X3) -- (X4);
    \draw[dashed,draw=black!66] (Y11) -- (Y22);
    \draw[dashed,draw=black!66] (Y11) -- (Y23);
    \draw[dashed,draw=black!66] (Y21) -- (Y31);
    \draw[dashed,draw=black!66] (Y23) -- (Y31);
    \draw[dashed,draw=black!66] (Y31) -- (Y42);

\end{tikzpicture}

\end{figure}

Each item set contains an $X$-item that corresponds to the satisfying assignment of $x$-variables, and $Y$-items that correspond to all possible unsatisfying assignments of $y$-variables, as well as special items. Forbidden pairs between items from different sets are marked with dashed lines (for example, $Y$-item from item set 1 cannot be taken simultaneously with two first $Y$-items from item set 4, as the former corresponds to $y_1=0$, while the latter to $y_1=1$, etc.). The solution in which $X$-items are selected from first 3 sets, and the special item from the set 4, has maximum regret equal to $3B+1$, since the adversary is forced to select an $X$-item from one of the first 3 sets.

\end{example}

\section{Solution Algorithm}\label{sec:sol-alg}

A standard technique for solving a general class of min-max regret optimization problems, for which the deterministic (nominal) version can be written as integer linear program, is based on solving a sequence of relaxed problems, with successively generated cuts in order to tighten the relaxation (see e.g., \cite{aissi2009min}). We develop a solution algorithm for \textsc{IRIS} based on this approach. Note that the worst-case cost scenario $\bb{c}(\bb{x})$ is an extreme point in $\mathcal{U}$, defined as: $$ c_{ij}(\bb{x}) = \left\{ 
\begin{array}{cc}
    c_{ij}^+, & x_{ij} = 1, \\
    c_{ij}^-, & x_{ij} = 0. 
\end{array}
\right., $$ which can be written as $c_{ij}(\bb{x}) = c_{ij}^- - (c_{ij}^- - c_{ij}^+)x_{ij}$. Using this fact we can rewrite the problem \eqref{generic-obj} as $$ \min_{\bb{x} \in \mathcal{F}} \left( \bb{c}(\bb{x}) \cdot \bb{x} - \min_{\bb{y} \in \mathcal{F}} \bb{c}(\bb{x}) \cdot \bb{y} \right), $$ where $\mathcal{F}$ is as defined in \eqref{F-constr}, and further: 
\begin{equation}\label{master-obj} 
    \min_{{\bb{x} \in \mathcal{F}, z \geq 0}} \left( \sum_{i=1}\sum_{j=1}^{r_i} c_{ij}^+ x_{ij} - z \right) 
\end{equation}
subject to: 
\begin{equation}\label{master-constr} 
    \forall_{\bb{y} \in \mathcal{C}} \;\; \bb{c}({\bb{x}}) \cdot \bb{y} \geq z, 
\end{equation}
where $\mathcal{C} = \mathcal{F}$ is the set of all feasible solutions of the \textsc{RIS} problem \eqref{F-constr}.

The above problem has exponentially many constraints (excluding trivial cases). Instead of solving it directly, we relax \eqref{master-constr} by selecting a subset $\mathcal{C} \subset \mathcal{F}$ of small size. After solving relaxed problem we obtain a solution $(\hat{\bb{x}}, \hat{z})$. Vector $\hat{\bb{x}}$ is a feasible solution of \eqref{generic-obj}, but not necessarily optimal. We can evaluate the maximum regret \eqref{max-regret} for $\hat{\bb{x}}$, and use the value $\hat{z}$ to check if $\hat{\bb{x}}$ is also feasible for the non-relaxed problem, by testing if
$$
    \sum_{i=1}\sum_{j=1}^{r_i} c_{ij}^+ \hat{x}_{ij} - \hat{z} \geq R(\hat{\bb{x}}).
$$
If this is the case, then $\hat{\bb{x}}$ must be optimal for \eqref{generic-obj}. Otherwise, the computed maximum regret $R(\hat{\bb{x}})$ produces an alternative solution $\hat{\bb{y}}$, which can be added to the set $\mathcal{C}$, tightening the relaxation. We resolve the new relaxed problem, and repeat the above procedure until either an optimal solution is found, or the stopping condition is met. The overview of this solution method is presented as Algorithm \ref{alg-1}.

\bigskip
\bigskip
\bigskip

\begin{algorithm}
	\caption{Solving decomposition by cut generation.}\label{alg-1}
\begin{algorithmic}[1]
    \State Initialize the set of cuts $\mathcal{C}$. Set small constant, e.g., $\varepsilon = 10^{-5}$.
    
    \State Solve the relaxed problem \eqref{master-obj}--\eqref{master-constr} for current $\mathcal{C}$, obtaining $(\hat{\bb{x}}, \hat{z})$. \label{alg:solve-master}
    
    \State Compute $\hat{\bb{y}}$ and $R(\hat{\bb{x}})$ by solving the \textsc{RIS} problem in scenario $\bb{c}(\hat{\bb{x}})$. \label{alg:solve-slave}
    
    \If {$$\sum_{i=1}\sum_{j=1}^{r_i} c_{ij}^+ \hat{x}_{ij} - \hat{z} \geq R(\hat{\bb{x}}) - \varepsilon $$}
         \State Return $\bb{x}^* = \hat{\bb{x}}$ and terminate the algorithm.
    \Else    
        \State Add cut $\hat{\bb{y}}$ to set $\mathcal{C}$, and go to step \ref{alg:solve-master}.
    \EndIf
\end{algorithmic}
\end{algorithm}

{\em Remarks.}

\begin{enumerate}
    \item Note that each time we solve the relaxed problem \eqref{master-obj}--\eqref{master-constr} we establish an increasingly better lower bound value $LB$ on \eqref{generic-obj} (by taking the optimal value of the objective function \eqref{master-obj}). We may also discover subsequently better upper bound values $UB$, since each $\hat{\bb{x}}$ is feasible for \eqref{generic-obj}. Thus the above method allows to determine the relative gap $g = (UB - LB) / UB$ to the optimal solution value. This is very useful in assessing the quality of suboptimal solutions.
    
    \item For large problem instances, the number of generated cuts $|\mathcal{C}|$ may become very large in later iterations of the algorithm. This may significantly slow down the algorithm. However, many of these cuts would be inactive for optimal solution. Consequently, we set the time limit (e.g., 1 minute) for solving the relaxed problem \eqref{master-obj}--\eqref{master-constr} in step \ref{alg:solve-master}, and if this limit is exceeded before an optimal solution to relaxation was found, we remove from $\mathcal{C}$ few constraints with the largest linear slacks in the previous solution, and retry solving. This assures that the size of the relaxed problem remains within reasonable range. Note, however, that in this process some constraints that would become essential at later iterations might be removed; they would need to be generated again and added to $\mathcal{C}$, and this may potentially consume additional computational time.
    
    \item When solving the relaxed problem \eqref{master-obj}--\eqref{master-constr} using branch and bound scheme, we may obtain a number of feasible solutions that can be used for the warm-start initialization when resolving with a new cut added. 
    
    \item For ``transitive'' instances (as in the second case of Theorem \ref{thm:2}), solving the auxiliary \textsc{RIS} subproblem in step \ref{alg:solve-slave} can be performed using a dedicated polynomial time algorithm for min-cost max-flow \cite{ahuja1988network}. Otherwise the resulting integer linear problem can be solved via general branch \& bound.
\end{enumerate}

\subsection{Initializing the set of cuts}

The performance of the algorithm described above depends on the choice of initial set of vectors $\mathcal{C} = \{ \bb{y}^{(k)} \}$ in Step 1. A simplest way to initialize it is to select a number of extreme scenarios $\bb{c}^{(k)}$ (vertices of $\mathcal{U}$), and obtain vectors $\bb{y}^{(k)}$ by solving deterministic \textsc{RIS} problem for these scenarios.

We initialize this set by randomly sampling 100 extreme scenarios and solving \textsc{RIS} problem for them. We also add to this set the set of solutions found using an evolutionary heuristic method, and use their best value as an initial upper bound $UB$ in the main algorithm. 

In the initialization stage, we first compute the mid-point scenario solution (it cannot be the worst-case, but often achieves a good solution value), and use it as an initial solution in an evolutionary heuristic. The latter method iteratively searches for improved solutions by applying mutation and crossover operations to the pool of current best solutions (population). Mutation operation randomly changes the selection of items within each set (the number of changes, as well as the number of affected sets, is also selected randomly). The crossover operation takes a pair of solutions, and replaces randomly selected sets from the first solution with the corresponding sets from the second solution. Each change is performed only if it results in a feasible solution. Both operations are repeated a fixed number of times, resulting in a new population of currently best solutions, which are passed to the next iteration. The final population is returned as a result. In the experiments we used 20 iterations on a population of size 10, applying 100 crossovers and mutations in each iteration.

\section{Experimental Results}

In this section we present the results of computational experiments conducted on a sequence of randomly generated problem instances. Each experiment consists of running the relaxation-based algorithm (with the initial set of cuts generated by the heuristic algorithm described in the previous section) on 10 instances for each fixed set of parameters. The considered parameters of \textsc{IRIS} problem instances are denoted: $m$ -- the number of item sets, $r_i$ -- the number of items in set $i$, $p_i$ -- the number of items required to be selected from set $i$, $K$ -- the number of constraints of type \eqref{ris-2}. Column labeled $n$ contains the number of (binary) decision variables in the resulting MIP. Bounds of cost intervals of all items were randomly generated integers between 1 and 100. All computations have been performed using CPLEX 12.8 optimization software and Python programming language.

We distinguish two main types of instances: (a) normal instances and (b) transitive-constraints instances. For the former, the constraints \eqref{ris-2} are generated by randomly sampling $K$ pairs of items, each from different set. For the latter, additionally, a transitive closure of the ``forbiddance'' relation is generated, and the additional constraints are added to restrict the selection of items to one per equivalence class. In the results, an average number of these constraints is reported in the column labeled ``K''.

Each row in Tables \ref{tab-1}--\ref{tab-2} contains the results from running the algorithm on 10 instances. First, the instance parameters are given, followed by the mean value and standard deviation of elapsed time for the algorithm to terminate. The stopping condition was that either an optimal solution has been found, or iteration limit of 500 has been reached. The next two columns contain the mean value and standard deviation of the iteration count of these runs that resulted in optimal solution. Finally, the number of instances (out of 10) solved to optimality is reported in the column ``opt.'', followed by the the mean solution value. For the suboptimal ones, the last column contains the mean relative optimality gap, i.e., the value $g = (UB - LB)/UB$.
\begin{table*}[ht]
    {\scriptsize
    \centering 
    \caption{Results for transitive-constraints instances for increasing average number of constraints.} \label{tab-1} 
    \begin{tabular}
        {|ccccc|cc|cc|ccc|} \hline \multicolumn{5}{|c|}{instance} & \multicolumn{2}{c|}{time (sec.)} & \multicolumn{2}{c|}{iterations} & \multicolumn{3}{c|}{solution value} \\
        $n$ & $m$ & $r_i$ & $p_i$ & $K$ & mean & std & mean & std & opt. & mean & gap \\
        \hline 50 & 5 & 10 & 2 & 5.90 & 3.63 & 1.74 & 17.60 & 14.02 & 10 & 240.20 & - \\
        50 & 5 & 10 & 2 & 13.50 & 9.65 & 17.55 & 18.90 & 10.21 & 10 & 242.10 & - \\
        50 & 5 & 10 & 2 & 28.20 & 4.08 & 1.78 & 22.80 & 16.30 & 10 & 236.50 & - \\
        50 & 5 & 10 & 2 & 63.20 & 2.94 & 0.68 & 12.60 & 7.12 & 10 & 212.40 & - \\
        50 & 5 & 10 & 2 & 108.50 & 3.73 & 2.16 & 19.50 & 19.68 & 10 & 236.90 & - \\
        50 & 5 & 10 & 2 & 216.40 & 3.30 & 0.90 & 15.80 & 10.27 & 10 & 245.30 & - \\
        \hline 50 & 5 & 10 & 3 & 5.90 & 76.49 & 86.88 & 182.89 & 102.15 & 9 & 339.90 & 0.03\\
        50 & 5 & 10 & 3 & 16.10 & 154.57 & 226.80 & 156.00 & 98.36 & 7 & 352.70 & 0.02\\
        50 & 5 & 10 & 3 & 27.40 & 83.21 & 203.43 & 82.56 & 42.70 & 9 & 339.10 & 0.02\\
        50 & 5 & 10 & 3 & 52.50 & 12.52 & 9.65 & 73.00 & 46.80 & 10 & 307.30 & - \\
        50 & 5 & 10 & 3 & 108.90 & 12.67 & 17.40 & 72.60 & 85.21 & 10 & 292.00 & - \\
        50 & 5 & 10 & 3 & 209.40 & 2.75 & 0.90 & 7.60 & 10.33 & 10 & 179.70 & - \\
        \hline 50 & 5 & 10 & 5 & 5.90 & 466.20 & 461.35 & 177.75 & 78.59 & 4 & 417.60 & 0.05 \\
        50 & 5 & 10 & 5 & 13.70 & 186.94 & 270.93 & 161.50 & 94.68 & 8 & 391.00 & 0.04 \\
        50 & 5 & 10 & 5 & 30.30 & 66.89 & 121.92 & 89.78 & 141.37 & 9 & 321.80 & 0.00\\
        50 & 5 & 10 & 5 & 60.00 & 2.81 & 0.93 & 11.80 & 10.71 & 10 & 231.10 & - \\
        50 & 5 & 10 & 5 & 89.50 & 2.11 & 0.19 & 2.20 & 1.83 & 10 & 167.10 & - \\
        \hline 100 & 10 & 10 & 2 & 5.30 & 332.93 & 275.29 & 265.44 & 103.93 & 9 & 427.80 & 0.01\\
        100 & 10 & 10 & 2 & 12.10 & 392.38 & 331.70 & 218.43 & 70.48 & 7 & 445.40 & 0.02\\
        100 & 10 & 10 & 2 & 18.90 & 390.49 & 457.45 & 237.89 & 152.66 & 9 & 445.50 & 0.01\\
        100 & 10 & 10 & 2 & 31.10 & 239.29 & 356.14 & 191.00 & 131.26 & 9 & 431.10 & 0.02\\
        100 & 10 & 10 & 2 & 43.70 & 77.80 & 62.76 & 144.70 & 69.12 & 10 & 418.70 & - \\
        100 & 10 & 10 & 2 & 63.10 & 346.88 & 376.60 & 216.25 & 131.68 & 8 & 460.70 & 0.02\\
        \hline 100 & 5 & 20 & 5 & 5.80 & 4658.68 & 2476.08 & 218.00 & 0.00 & 1 & 545.50 & 0.07\\
        100 & 5 & 20 & 15 & 5.80 & 3536.08 & 2414.70 & 219.00 & 118.43 & 4 & 500.30 & 0.03\\
        100 & 10 & 10 & 5 & 5.30 & 7756.25 & 2332.56 & - & - & 0 & 789.00 & 0.17\\
        100 & 10 & 10 & 5 & 12.10 & 7721.47 & 2311.46 & - & - & 0 & 788.40 & 0.16\\
        150 & 5 & 30 & 5 & 5.20 & 5690.23 & 3108.90 & 338.00 & 107.24 & 3 & 527.90 & 0.06\\
        150 & 5 & 30 & 25 & 5.10 & 3247.63 & 3843.27 & 147.57 & 78.99 & 7 & 469.50 & 0.04\\
        \hline
    \end{tabular}
    }
\end{table*}
\begin{table*}[ht] 
    {\scriptsize
    \centering 
    \caption{Results for normal instances.} \label{tab-2} 
    \begin{tabular}
        {|ccccc|cc|cc|ccc|} \hline \multicolumn{5}{|c|}{instance} & \multicolumn{2}{c|}{time (sec.)} & \multicolumn{2}{c|}{iterations} & \multicolumn{3}{c|}{solution value} \\
        
        $n$ & $m$ & $r_i$ & $p_i$ & $K$ & mean & std & mean & std & opt. & mean & gap \\
        
        \hline 50 & 5 & 10 & 3 & 5 & 69.57 & 86.86 & 167.78 & 91.93 & 9 & 336.80 & 0.03\\
        50 & 5 & 10 & 3 & 10 & 40.42 & 30.76 & 168.30 & 85.81 & 10 & 335.70 & - \\
        50 & 5 & 10 & 3 & 15 & 20.82 & 20.11 & 100.50 & 67.02 & 10 & 318.50 & - \\
        50 & 5 & 10 & 3 & 20 & 22.61 & 26.27 & 95.60 & 94.13 & 10 & 306.20 & - \\
        50 & 5 & 10 & 3 & 25 & 9.94 & 4.12 & 64.00 & 29.70 & 10 & 298.10 & - \\
        50 & 5 & 10 & 3 & 30 & 12.55 & 19.72 & 62.00 & 76.40 & 10 & 316.10 & - \\
        50 & 5 & 10 & 3 & 50 & 9.13 & 5.82 & 54.90 & 33.02 & 10 & 338.50 & - \\
        50 & 5 & 10 & 3 & 75 & 3.96 & 1.75 & 15.40 & 11.98 & 10 & 311.00 & - \\
        50 & 5 & 10 & 3 & 100 & 4.83 & 3.53 & 15.70 & 19.92 & 10 & 294.30 & - \\
        
        \hline 50 & 5 & 10 & 5 & 5 & 415.87 & 408.31 & 225.00 & 117.13 & 5 & 414.90 & 0.05\\
        50 & 5 & 10 & 5 & 10 & 76.25 & 75.29 & 207.80 & 138.24 & 10 & 374.00 & - \\
        50 & 5 & 10 & 5 & 15 & 111.64 & 141.01 & 157.50 & 101.02 & 8 & 384.50 & 0.01\\
        50 & 5 & 10 & 5 & 20 & 76.98 & 119.22 & 127.89 & 124.26 & 9 & 402.10 & 0.01\\
        50 & 5 & 10 & 5 & 25 & 42.38 & 92.56 & 107.20 & 128.27 & 10 & 371.40 & - \\
        50 & 5 & 10 & 5 & 30 & 7.64 & 3.16 & 45.20 & 17.74 & 10 & 364.70 & - \\
        50 & 5 & 10 & 5 & 50 & 3.76 & 1.11 & 10.70 & 8.84 & 10 & 328.50 & - \\
        \hline 150 & 5 & 30 & 5 & 10 & 5904.29 & 2812.44 & 212.00 & 0.00 & 1 & 532.30 & 0.03\\
        150 & 5 & 30 & 25 & 10 & 1721.59 & 2501.74 & 142.50 & 83.83 & 8 & 467.10  & 0.02\\
        100 & 10 & 10 & 5 & 10 & 7968.38 & 2066.25 & - & - & 0 & 797.00 & 0.17\\
        150 & 15 & 10 & 5 & 10 & 9009.75 & 1127.45 & - & - & 0 & 1217.20 & 0.31\\
        200 & 20 & 10 & 5 & 10 & 6153.30 & 1217.37 & - & - & 0 & 1622.60 & 0.37\\
        \hline 
    \end{tabular}
    }
\end{table*}

We observe that the hardest instances often contain only few constraints of type \eqref{ris-2}. This is due to the fact that the larger the number $K$, the smaller is the set of feasible solutions $\mathcal{F}$, making it easier to explore. On the other hand, they are typically difficult when it is required to select about the half of the items in each set.

\section{Conclusions}

We presented a generalization of the representatives selection problem with interval costs uncertainty, and considered it in the robust optimization framework with the maximum regret criterion. We characterized the computational complexity of the problem, and developed both exact and heuristic solution method. The main solution algorithm consists of cut generation and solving a sequence of relaxed problems. An evolutionary heuristic stage was proposed to initialize the algorithm with a set of cuts. While the considered robust problem is difficult to solve to optimality, we have demonstrated that it is possible to solve it for moderately-sized problem instances, and approximate solutions using a combination of heuristics and exact method. 

The analysis of this problem can be extended in the future for different types of restrictions on item choices. For instance, some items may be required to be selected in bundles, and would be useless without simultaneously selecting specific items from other sets.

%\bibliographystyle{spmpsci}      % mathematics and physical sciences%\bibliographystyle{spmpsci}      % mathematics and physical sciences
%\bibliographystyle{spphys}       % APS-like style for physics%\bibliographystyle{spphys}       % APS-like style for physics
%\bibliography{MD_arxiv} % \bibliographystyle{plain}
% \bibliography{main}   % name your BibTeX data base

\begin{thebibliography}{10}
\providecommand{\url}[1]{{#1}}
\providecommand{\urlprefix}{URL }
\expandafter\ifx\csname urlstyle\endcsname\relax
  \providecommand{\doi}[1]{DOI~\discretionary{}{}{}#1}\else
  \providecommand{\doi}{DOI~\discretionary{}{}{}\begingroup
  \urlstyle{rm}\Url}\fi

\bibitem{ahuja1988network}
Ahuja, R.K., Magnanti, T.L., Orlin, J.B.: Network flows.
\newblock Prentice Hall (1988)

\bibitem{aissi2005complexity}
Aissi, H., Bazgan, C., Vanderpooten, D.: Complexity of the min--max and
  min--max regret assignment problems.
\newblock Operations Research Letters \textbf{33}(6), 634--640 (2005)

\bibitem{aissi2009min}
Aissi, H., Bazgan, C., Vanderpooten, D.: Min--max and min--max regret versions
  of combinatorial optimization problems: A survey.
\newblock European Journal of Operational Research \textbf{197}(2), 427--438
  (2009)

\bibitem{averbakh2001complexity}
Averbakh, I.: On the complexity of a class of combinatorial optimization
  problems with uncertainty.
\newblock Mathematical Programming \textbf{90}(2), 263--272 (2001)

\bibitem{averbakh2004interval}
Averbakh, I., Lebedev, V.: Interval data minmax regret network optimization
  problems.
\newblock Discrete Applied Mathematics \textbf{138}(3), 289--301 (2004)

\bibitem{bertsimas2003robust}
Bertsimas, D., Sim, M.: Robust discrete optimization and network flows.
\newblock Mathematical programming \textbf{98}(1-3), 49--71 (2003)

\bibitem{conde2004improved}
Conde, E.: An improved algorithm for selecting p items with uncertain returns
  according to the minmax-regret criterion.
\newblock Mathematical Programming \textbf{100}(2), 345--353 (2004)

\bibitem{deineko2013complexity}
Deineko, V.G., Woeginger, G.J.: Complexity and in-approximability of a
  selection problem in robust optimization.
\newblock 4OR \textbf{11}(3), 249--252 (2013)

\bibitem{dolgui2012min}
Dolgui, A., Kovalev, S.: Min--max and min--max (relative) regret approaches to
  representatives selection problem.
\newblock 4OR \textbf{10}(2), 181--192 (2012)

\bibitem{drwal2018robust}
Drwal, M.: Robust scheduling to minimize the weighted number of late jobs with
  interval due-date uncertainty.
\newblock Computers \& Operations Research \textbf{91}, 13--20 (2018)

\bibitem{drwal2016complexity}
Drwal, M., Rischke, R.: Complexity of interval minmax regret scheduling on
  parallel identical machines with total completion time criterion.
\newblock Operations Research Letters \textbf{44}(3) (2016)

\bibitem{garey2002computers}
Garey, M., Johnson, D.: Computers and Intractability: A Guide to the Theory of
  NP-completeness.
\newblock W.H. Freeman (2002)

\bibitem{goerigk2016algorithm}
Goerigk, M., Sch{\"o}bel, A.: Algorithm engineering in robust optimization.
\newblock In: Algorithm engineering, pp. 245--279. Springer (2016)

\bibitem{JAHROMI2012224}
Jahromi, M., Tavakkoli-Moghaddam, R.: A novel 0-1 linear integer programming
  model for dynamic machine-tool selection and operation allocation in a
  flexible manufacturing system.
\newblock Journal of Manufacturing Systems \textbf{31}(2), 224 -- 231 (2012)

\bibitem{kall1994stochastic}
Kall, P., Wallace, S.W., Kall, P.: Stochastic programming.
\newblock Springer (1994)

\bibitem{kasperski2008discrete}
Kasperski, A.: Discrete optimization with interval data: minmax regret and
  fuzzy approach.
\newblock Springer (2008)

\bibitem{kasperski2015approximability}
Kasperski, A., Kurpisz, A., Zieli{\'n}ski, P.: Approximability of the robust
  representatives selection problem.
\newblock Operations Research Letters \textbf{43}(1), 16--19 (2015)

\bibitem{kasperskiminmax}
Kasperski, A., Zielinski, P.: Minmax (regret) scheduling problems.
\newblock Sequencing and Scheduling with Inaccurate Data. Y. Sotskov F. Werner
  (eds.) pp. 159--210 (2014)

\bibitem{LEE200361}
Lee, C.S., Kim, S.S., Choi, J.S.: Operation sequence and tool selection in
  flexible manufacturing systems under dynamic tool allocation.
\newblock Computers \& Industrial Engineering \textbf{45}(1), 61 -- 73 (2003)

\bibitem{pereira2011exact}
Pereira, J., Averbakh, I.: Exact and heuristic algorithms for the interval data
  robust assignment problem.
\newblock Computers \& Operations Research \textbf{38}(8), 1153--1163 (2011)

\bibitem{roy2010robustness}
Roy, B.: Robustness in operational research and decision aiding: A
  multi-faceted issue.
\newblock European Journal of Operational Research \textbf{200}(3), 629--638
  (2010)

\bibitem{schrijver1998theory}
Schrijver, A.: Theory of linear and integer programming.
\newblock John Wiley \& Sons (1998)

\bibitem{stockmeyer1976polynomial}
Stockmeyer, L.J.: The polynomial-time hierarchy.
\newblock Theoretical Computer Science \textbf{3}(1), 1--22 (1976)

\end{thebibliography}

\end{document}